\documentclass[aps,prl,twocolumn,amsmath,amssymb,nofootinbib,10pt]{revtex4-1}
\usepackage{pslatex,graphicx,dcolumn,bm,natbib}
\usepackage{bm}        
\usepackage{mathrsfs} 
\usepackage{color}
\newcommand{\de}{\Delta \varepsilon}
\newcommand{\mde}{\overline{\Delta \varepsilon}}
\newcommand{\dpp}{\lvert p_0-p_0^*\rvert}
\usepackage[normalem]{ulem}

\begin{document}
	
	\author
	{
	Dapeng Bi$^1$, J. H. Lopez$^1$, J. M. Schwarz$^{1,2}$, and M. Lisa Manning$^{1,2}$\\
		\normalsize{
		$^1$Department of Physics, Syracuse University, Syracuse, NY 13244, USA \\
		$^2$Syracuse Biomaterials Institute, Syracuse, NY 13244, USA \\
		}
	}

\title{A density-independent rigidity transition in biological tissues}

\begin{abstract} %
Cell migration is important in many biological processes, including embryonic development, cancer metastasis, and wound healing.  In these tissues, a cell's motion is often strongly constrained by its neighbors, leading to glassy dynamics. 
While self-propelled particle models exhibit a density-driven glass transition, this does not explain liquid-to-solid transitions in confluent tissues, where there are no gaps between cells and therefore the density is constant.  Here we demonstrate the existence of a new type of rigidity transition that occurs in the well-studied vertex model for confluent tissue monolayers at constant density.  We find the onset of rigidity is governed by a model parameter that encodes single-cell properties such as cell-cell adhesion and cortical tension, providing an explanation for a liquid-to-solid transitions in confluent tissues and making testable predictions about how these transitions differ from those in particulate matter.
\end{abstract}

\maketitle

Important biological processes such as embryogensis, tumorigenesis, and wound healing require cells to move collectively within a tissue. Recent experiments suggest that when cells are packed ever more densely, they start to exhibit collective motion~\cite{Schoetz2013,angelini_2011,KaesNJP} traditionally seen in non-living disordered systems such as colloids, granular matter or foams~\cite{haxton_liu, abate_durian, LiuNagelReview}. These collective behaviors exhibit growing timescales and lengthscales associated with rigidity transitions. 

Many of these effects are also seen in Self-Propelled Particle (SPP) models~\cite{Vicsek_1995}. In SPP models, overdamped particles experience an active force that causes them to move at a constant speed, and particles change direction due to interactions with their neighbors or an external bath. To model cells with  a cortical network of actomyosin and adhesive molecules on their surfaces, particles interact as repulsive disks or spheres, sometimes with an additional short-range attraction~\cite{Henkes2011,Chate}. These models generically exhibit a glass transition at a critical packing density of particles, $\phi_c$, where $\phi_c < 1$~\cite{Henkes2011, Berthier2014, Berthier_Kurchan, Schoetz2013}, and near the transition point they exhibit collective motion~\cite{Henkes2011} that is very similar to that seen in experiments~\cite{Silberzan2010}. 
 
An important open question is whether the density-driven glass transition in SPP models explains the glassy behavior observed in non-proliferating confluent biological tissues, where there are no gaps between cells and the packing fraction $\phi$ is fixed at precisely unity.  For example, zebrafish embryonic explants are confluent three-dimensional tissues where the cells divide slowly and therefore the number of cells per unit volume remains nearly constant.  Nevertheless, these tissues exhibit hallmarks of glassy dynamics such as caging behavior and viscoelasticity.  Furthermore, ectoderm tissues have longer relaxation timescales than mesendoderm tissues, suggesting ectoderm tissues are closer to a glass transition, despite the fact that both tissue types have the same density~\cite{Schoetz2013}. This indicates that there should be an additional parameter controlling glass transitions in confluent tissues.

In this work, we study confluent monolayers using the vertex model
\cite{ngai_honda,Farhadifar2007, Hufnagel2007, Staple2010, Hilgenfeldt2008,manning_2010, Wang2012, ChiouShraiman,vertex_model_review}, 
to determine how tissue mechanical response varies with single-cell properties such as adhesion and cortical tension. 
We find a new type of rigidity transition that is not controlled by the density, but instead by a dimensionless \textit{target shape index} that is specified by single-cell properties.   
This rigidity transition possesses several hallmarks of a second-order phase transition.
These findings provide a novel explanation for liquid-to-solid transitions in tissues that remain at constant density.

The vertex model, which agrees remarkably well with experimental data for confluent monolayers  
~\cite{ngai_honda,Farhadifar2007, Hufnagel2007, Staple2010, Hilgenfeldt2008,manning_2010, Wang2012, ChiouShraiman,vertex_model_review}, approximates the monolayer as a collection of adjacent columnar cells. The mechanical energy of a single cell labeled `$i$' is given by~\cite{Farhadifar2007, Staple2010}:
\begin{equation}
\label{single_cell_energy}
\quad E_i = \beta_{i} (A_i-A_{i0})^2 + \xi_i P_i^2 + \gamma_i P_i.
\end{equation}
 The first term results from a combination of 3D cell incompressibility and the monolayer's resistance to height fluctuations or  cell bulk elasticity ~\cite{Hufnagel2007,Angelini_cell_volume}. Then $\beta_i$ is a \textit{height elasticity}, and $A_i$ and $A_{i0}$ are the actual and preferred cell cross-sectional areas. 

\begin{figure}[htpb]
\begin{center}
\includegraphics[width=1\columnwidth]{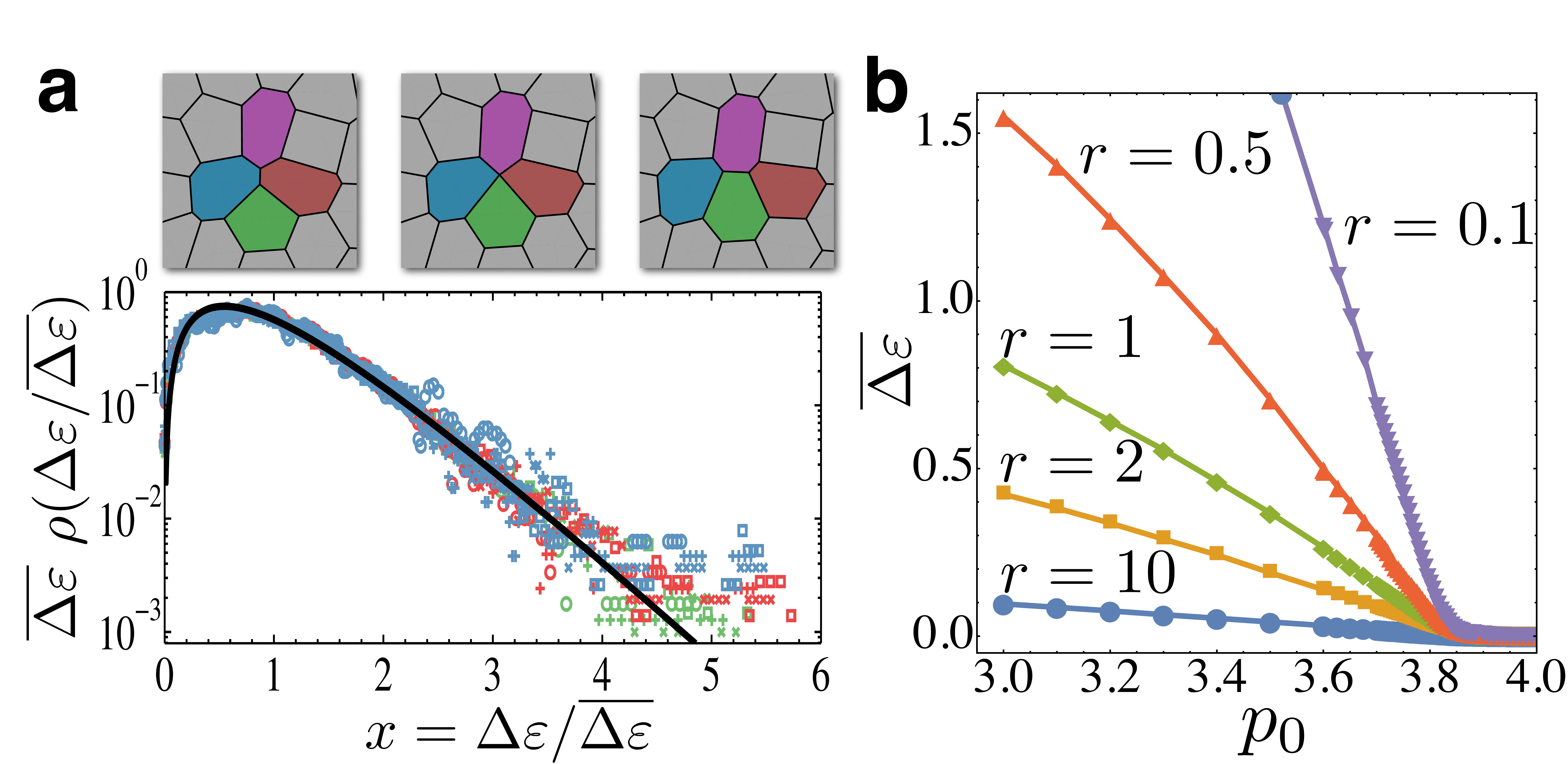}
\caption{
{\bf Energy barriers for local cellular rearrangements.} (a)
Illustration of a T1 transition in a confluent tissue 
and the normalized distribution $\rho$ of normalized energy barrier heights $\de/ \mde$ for a large range of parameters ($r = 0.5, 1, 2$   and $p_0 = 3.2 - 3.7$). They have a universal shape well-fit by a $k-gamma$ distribution (solid line), 
indicating that $\mde$ completely specifies the distribution and describe the mechanical . 
(b) $\mde$ as function of the target shape index $p_0$ for various values of the inverse perimeter modulus $r$. 
}
\label{de_stat}
\end{center}
\end{figure}
 
\begin{figure*}[t]
\centering
\includegraphics[width=2\columnwidth]{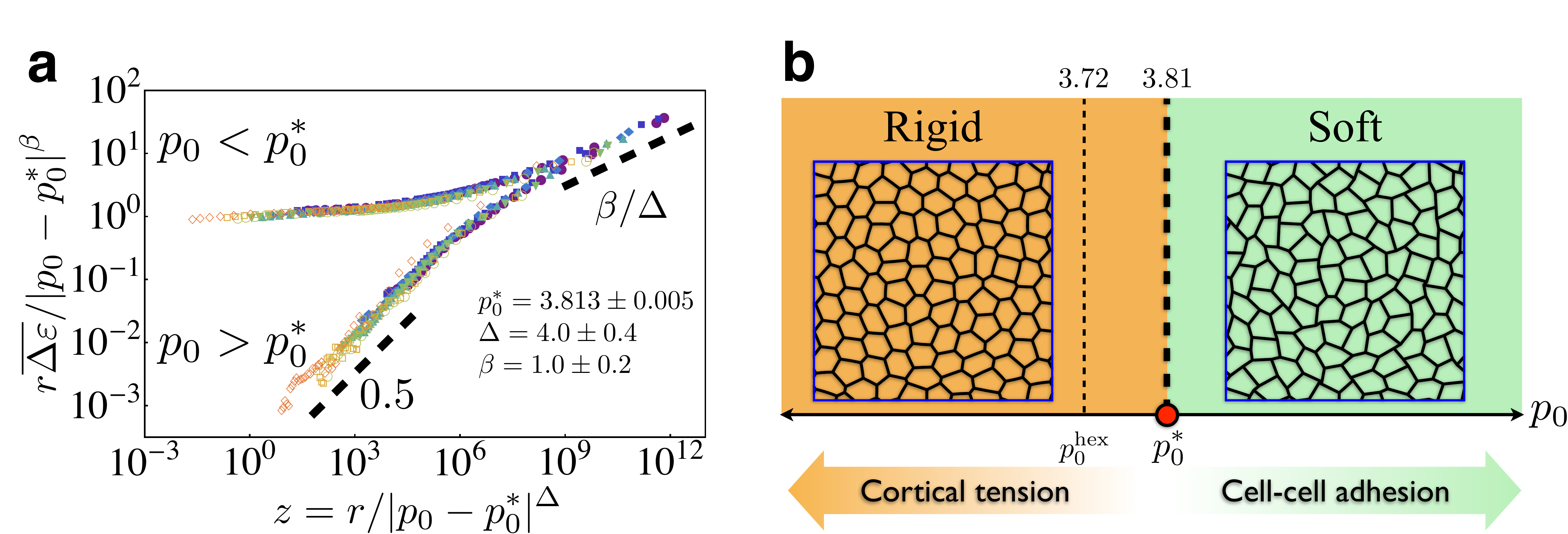}
\caption{
{\bf A rigidity transition in confluent tissues} 
(a) 
Critical scaling collapse of the average energy barrier height $\mde$, normalized by multiplying $r/\dpp^\beta$, as a function of $z=r  / \dpp^\Delta$  for the data shown in Fig.~\ref{de_stat}(b), confirming the scaling ansatz of equation~\eqref{crit_scaling}.
(b) 
The rigidity transition is demonstrated in a simple phase diagram as function of $p_0$, snapshots are taken from a typical rigid tissue ($p_0=3.7$) and soft tissue ($p_0=3.96$). A rigidity transition occurs at $p_0=p_0^*\approx 3.813$ for disordered metastable tissue configurations. 
The line corresponding to the order-to-disorder transition reported by Staple et al~\cite{Staple2010}  is shown for comparison. Below $p_0^{\text{hex}}$, the ground state is a hexagonal lattice and above $p_0^{\text{hex}}$ the ground state is disordered. 
}
\label{scaling_collapse}
\end{figure*}

The second term in equation~\eqref{single_cell_energy} 
is quadratic in the cell cross-sectional perimeter $P_i$ and
 models the active contractility of the actin-myosin subcellular cortex, with elastic constant $\xi_i$~\cite{Farhadifar2007}, and the last term represents an interfacial tension $\gamma_i$ set by a competition between the cortical tension and the energy of cell-cell adhesion~\cite{Graner1992, manning_2010} between two contacting cells. $\gamma_i$ can be positive if the cortical tension is greater than the adhesive energy, or negative if the adhesion dominates.  It is also possible to incorporate strong feedback between adhesion and cortical tension in this term~\cite{Amack2012,manning_2010}.
  Since only the effective forces -- the derivatives of the energy with respect to the degrees of freedom -- are physically relevant, equation~\eqref{single_cell_energy} can be rewritten: $E_i = \beta_{i} (A_i-A_{i0})^2 + \xi_i {(P_i - P_{i0})}^2$,
where $P_{i0} = -\gamma_i / (2 \xi_i)$ is an effective target shape index.

As discussed in~\cite{Staple2010}, when all single-cell properties are equal ($\beta_i = \beta$, $\xi_i=\xi$, $A_{i0}=A_0$, $P_{i0}=P_0$), the total mechanical energy of a tissue containing $N$ cells can be non-dimensionalized:
\begin{equation}
\label{scaled_e_tot}
\mathcal{\epsilon} =\frac{1}{\beta A_0^2} \sum_{i}^{N} E_i = \sum_{i} \left [ (\tilde{a}_i-1)^2 + \frac{(\tilde{p}_i - p_0)^2}{r} \right],
\end{equation}
where $\tilde{a}_i = A_i/A_0$ and $\tilde{p}_i = P_i /\sqrt{A_0}$  are the rescaled shape functions for area and perimeter.
$r=\beta A_0^2/ \xi $ is the \textit{inverse perimeter modulus}
and $p_0 = P_0/\sqrt{A_0}$ is the \textit{target shape index}~\cite{note_p0}
or a preferred perimeter-to-area ratio; geometrically,
a regular hexagon corresponds to $p_0^{hex}=2 \sqrt{2} \sqrt[4]{3}\approx3.72$ and a regular pentagon to $p_0^{pent}=2 \sqrt{5} {(5-2 \sqrt{5})}^{1/4}\approx3.81$. 
 While we focus on the simple case where cells are identical, the rigidity transition is robust to small variations in cell properties(see Supplementary Materials).

In non-biological materials, bulk quantities such as shear/bulk modulus, shear viscosity and yield stress are often used to describe the mechanical response to external perturbations. However, cells are self-propelled and even in the absence of external forces, cells in confluent tissues regularly intercalate, or exchange neighbors~\cite{Lecuit_review,Guillot_sci_review_2013}.
In an isotropic confluent tissue monolayer where mitosis (cell division) or apoptosis (cell death) are rare, cell neighbor exchange must happen through intercalation processes known as  T1 transitions~\cite{Weaire_book,bi_softmatter},  where an edge between two cells shrinks to a point and a new edge arises between two neighboring cells as illustrated in Fig.~\ref{de_stat} (a). 
The mechanical response of the tissue is governed by the rate of cell rearrangements, and within the vertex model, the rate of T1 rearrangements is related to the amount of mechanical energy required to execute a T1 transition~\cite{bi_softmatter}.  Therefore, we first study how these energy barriers change with single-cell properties encoded in the model parameters $r$ and $p_0$.

To explore the statistics of energy barriers, we test all possible T1 transition paths (see Methods section) in 10 randomly generated  disordered   samples each consisting of $N=64$ cells. For each value of $p_0$ and $r$ tested, we obtained the distribution of energy barrier heights $\rho(\de)$. The functional form of the distribution becomes universal (Fig.~\ref{de_stat} (a)) when scaled by the mean energy barrier height $\mde(r, p_0)$. The rescaled distribution is well-fit by a $k-gamma$ distribution ($k^k/(k-1)! \; x^{k-1} \exp(-kx)$) with $x=\de/\mde $ and $k=2.2 \pm 0.2$. The $k-gamma$ distribution has been observed in many non-biological systems disordered systems~\cite{Brujic, Aste,bi_epl}, and generically results from maximizing the entropy subject to constraints~\cite{Aste,bi_epl}.  This confirms that the distribution of energy barriers depends on the single-cell properties $p_0$ and $r$ only through its average $\mde$.

Fig.~\ref{de_stat}(b) shows the dependence of $\mde$ on $p_0$ for various values of $r$. At $p_0 \lesssim 3.8$, the energy barriers are always finite, i.e. cells must put in some amount of work in order to deform and rearrange.  Here the tissue behaves like a solid; it is a rigid material with a finite yield stress. 
As $p_0$ is increased, the energy barriers decrease and become vanishingly small in the vicinity of $p_0 \approx 3.8$, so that cell shape deformations require no energy. This change in the mechanical behavior as function of $p_0$ is suggestive of a rigidity transition.

To better understand the nature of this rigidity transition, we search for a scaling collapse.  We use the scaled energy barrier value $r \mde(r,p_0)$ as an order parameter, since $r$ controls the overall scale of $\mde$ away from the transition. From Fig.~\ref{de_stat}(b), we also see that $r$ controls the sharpness of the transition, playing a role similar to the magnetic field in the Ising model. This is reasonable because as seen in Eq.~\ref{scaled_e_tot}, $r$ controls the strength of fluctuations in the perimeter. 
Assuming that the mechanical rigidity of the tissue is controlled by a critical point at some $p_0=p_0^*$, then near the critical point the order parameter $r \mde$ should be related to the variable that controls the fluctuations $r$ by a universal  ansatz\cite{Kadanoff_scaling}:
\begin{equation}
r \ \mde = \dpp^\beta f_{\pm} \left( \frac{r}{\dpp^\Delta}  \right).
\label{crit_scaling}
\end{equation}
Here $z=r/\dpp^\Delta$ is the crossover scaling variable, $\Delta$ is the crossover scaling critical exponent, and $f_{-}, f_{+}$ are the two branches of the crossover scaling functions for $p_0<p_0^*$ and $p_0>p_0^*$, respectively. 

After re-plotting the data in Fig.~\ref{de_stat}(b) using  equation~\eqref{crit_scaling}, we find an excellent scaling collapse onto two branches with $\Delta=4.0\pm0.4$, $\beta = 1.0\pm0.2$ and a precise location of the critical point $p_0^*=3.813\pm0.005$ as shown in Fig.~\ref{scaling_collapse}.  
In the mechanically rigid branch ($p_0 < p_0^*$), as $z \to 0$, $f_{-}$ is finite, meaning that the energy barrier is finite and scales as $\mde \propto \left( p_0^*-p_0\right)^\beta / r$. 
At the critical point ($p_0=p_0^*$), the two branches of the scaling function merge and $f_{+} = f_{-}=z^{\beta / \Delta}$ resulting in the scaling $\mde \propto r^ {\beta / \Delta-1}$.
The mechanically soft or fluid branch ($p_0 > p_0^*$) decays as $z^{0.5}$ as $z \to 0$ or $\mde \propto r^{-0.5} /\left( p_0-p_0^*\right)$. 
This scaling collapse is similar to those seen in jamming in particulate matter~\cite{olsson_teitel,haxton_liu} and rigidity percolation on random networks~\cite{das_quint_schwarz_rigidity,chase_rigidity,frey}, suggesting that $p_0^*$ is a critical point analogous to Point $J$ in the jamming transition or the critical occupation probability $p^*$ in random network models. 
However, unlike the jamming transition which is  density driven, density can not control the rigidity transition in the vertex model because everything takes place at a packing fraction of unity. Instead, this model displays a novel rigidity transition controlled by the target shape index, $p_0$. Fig.~\ref{scaling_collapse}(b) summarizes these results by a simple phase diagram and depicts two snapshots from rigid and soft simulations.

Although we calculate T1 transitions by shortening or lengthening a single cell-cell contact, our analysis of these local perturbations suggest a critical mechanical response with a growing lengthscale. To confirm and quantify these changes in the macroscopic mechanical response, we study the vibrational spectrum of the dynamical matrix~\cite{Ashcroft, Silbert2005}.  
 
We diagonalize the dynamical matrix to obtain normal modes and their corresponding eigenvalues $\{ \lambda_i \}$  (Methods) and  eigenfrequencies $\omega_i=\text{sign}(\lambda_i) \sqrt{ \vert \lambda_i \vert}$. Besides the trivial global translation modes which have $\omega=0$, any other non-positive $\omega$ correspond to a soft mode.  
 
The cumulative density of states is defined as the cumulative distribution function of $\omega$, 
\begin{equation}
N(\omega) = \int_{-\infty}^{\omega} D(\omega') d\omega' 
\label{cdf_def}
\end{equation}
where $D(\omega)$ is the density of states.  
 
If $N(\omega=0) > 0$, there are floppy modes -- collective displacements of the vertices that cost zero energy -- and the system is a fluid, while if $N(\omega) \rightarrow 0$ as $\omega \rightarrow 0$, any linear combination of displacements costs finite energy and the material is a solid. 
 
\begin{figure}[htpb]
\begin{center}
\includegraphics[width=1\columnwidth]{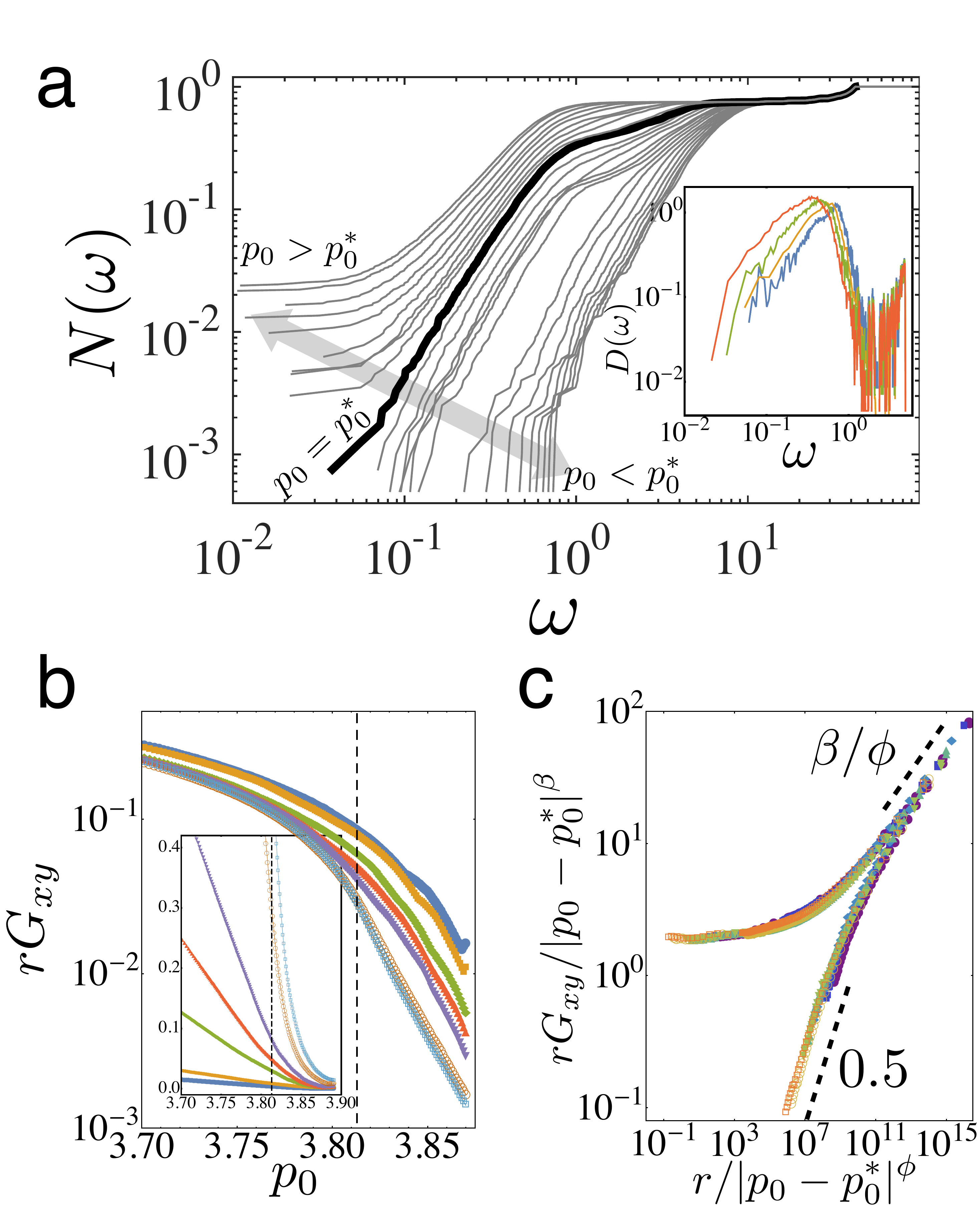}
\caption{
{\bf Analysis of mechanical properties.} 
(a)
The cumulative vibrational density of states $N(\omega)$ exhibits a rigidity transition at $r=0.1$ and $p_0*=3.813$ (thick line). Thin lines correspond to $r=0.1$ and values of $p_0$ ranging from  $3.78$ to $3.83$ in increments of $10^{-3}$.   Inset: Vibrational density of states $D(\omega)$ for selected $p_0$ values, from left to right: 3.78, 3.79, 3.80, 3.81.  At low $\omega$, $D(\omega) \sim \omega$ follows Debye scaling before arriving at a Boson peak. As $p_0$ is decreased toward the rigidity transition, the Boson peak also shifts to lower frequencies.
(b) 
The scaled shear modulus, $r G_{xy}$, as a function of $r$ (top to bottom: $r=0.05,0.1,0.5,1,2,10,20$) and $p_0$. Inset: linear scale of the same plot.
(c) Scaling of $G_{xy}$ near the rigidity transition obeys $r G_{xy} \dpp^{-\beta} = g_{\pm}(r \dpp^{-\phi})$, where $\beta=1.0\pm0.2$ and $\phi=5\pm+0.5$. 
}
\label{vibration_shear}
\end{center}
\end{figure}

Fig.~\ref{vibration_shear}(a)  demonstrates that the collective linear response exhibits a rigidity transition at $p_0 = p_0^*=3.813$, which is identical to the transition identified by our local, nonlinear energy barrier analysis. For $p_0<p_0^*$, $N(\omega)$ exhibits Debye scaling and approaches zero at zero frequency ($N(\omega) \sim \omega^d =\omega^2$), while for $p_0>p_0^*$,  $N$ exhibits a finite plateau at the lowest frequencies.
In addition, as the system approaches the rigidity transition from the solid phase, the density of states $D(\omega)$ exhibits a peak that shifts to lower frequencies (Fig.~\ref{vibration_shear}(a)), just as the so-called Boson Peak~\cite{Wyart2005,Silbert2005,ManningEPL2015} in jammed particle packings and glasses.  Interestingly, the shape and scaling of the peak is different from those in particulate matter, and this is an interesting avenue for future research.

Another standard measure of linear mechanical response is the shear modulus. We probe the tissue near the rigidity transition in response to a quasistatic simple shear strain $\gamma_{xy}$ and calculate the shear modulus $G_{xy}$(Methods). The shear modulus behaves similarly to  $\mde$ as function $p_0$ and $r$. As the $p_0$ is increased towards $p_0^*$, $G_{xy}$ drops rapidly (Fig.~\ref{vibration_shear}(b)) at the rigidity transition, with $r$ controlling the overall magnitude and the sharpness of the transition. Therefore, we propose a similar scaling ansatz for $G_{xy}$: 
\begin{equation}
r G_{xy}=  \dpp^{\beta} g_{\pm}\left(\frac{r}{\dpp^{\phi}}\right),
\label{shear_modulus_scaling}
\end{equation}
similar to equation~\eqref{crit_scaling}.  Again,the data Fig.~\ref{vibration_shear}(c) collapses onto two branches, with scaling exponents $\beta=1.0\pm0.2 \ \& \ \phi=5.0+0.5$. The upper branch corresponds to rigid tissues with a finite shear modulus, i.e. $z <\!< 1$, $g_{+}(z) \to \text{const}$ or $G_{xy} \sim {(p_0^*-p_0)}^\beta/r$. The lower branch corresponds to soft tissues with vanishing shear modulus, i.e., $g_{-}(z) \to \sqrt{z}$ or $G_{xy} \sim {(p_0-p_0^*)}^{\beta-\phi}/\sqrt{r} $ as $z \to 0$. At the transition, $G_{xy}$ becomes independent of $\dpp$ and scales as $G_{xy} \sim r^{\beta/\phi}$.

 An obvious remaining question is what sets the critical point $p_0^* \sim 3.81$.
To answer this question, we first study a simple mean-field model for a T1 topological swap.  In an infinite confluent tissue, the topological Gauss-Bonnet theorem requires each cell to have six neighbors on average~\cite{Weaire_book}. Therefore our mean-field model consists of four adjacent six-sided cells. To mimic the effect of additional neighboring cells, we fix each cell area equal to unity. Equation~\eqref{scaled_e_tot} then becomes:
\begin{equation}
\varepsilon_4 = \sum_{4 \ cells}(\tilde{p}_i -p_0)^2; \quad a_i = 1.
\label{four_cell_energy}
\end{equation}

\begin{figure}[htbp]
\begin{center}
\includegraphics[width=1\columnwidth]{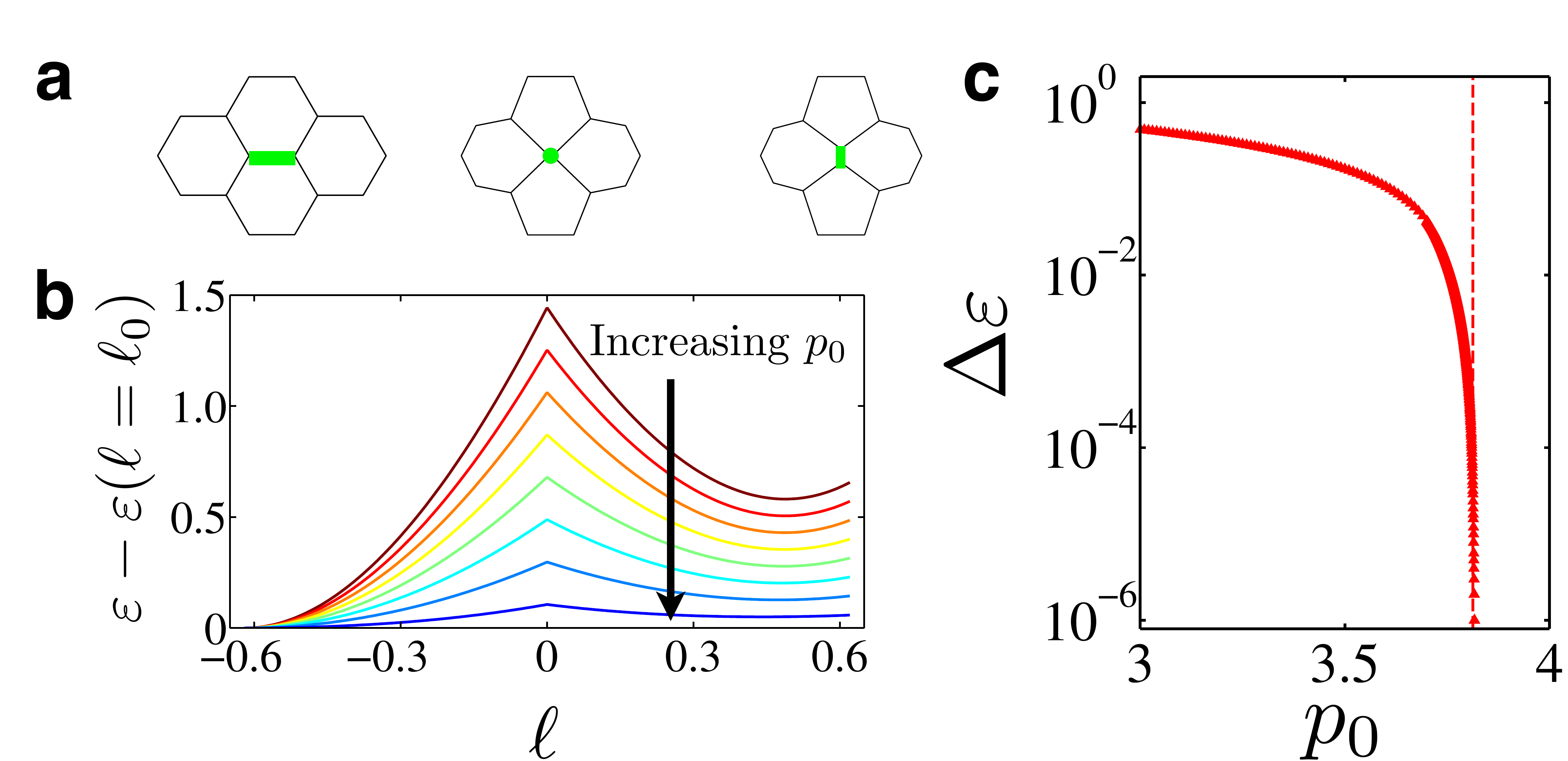}
\caption{
{\bf A simple four-cell model} 
(a) A four-cell aggregate undergoing a T1 topological swap. The thick (green) edge represents the cell-cell interface that is contracted to a point and then resolved in the perpendicular direction. (b) Energy of a four cell aggregate during a T1 transition, which attains a maximum at the transition point. $p_0$ varies from $1.5$ to $3.8$ in equal increments. 
(c) Energy barrier height as function of $p_0$ for a four-cell aggregate and a mean-field estimate (dashed) for the value of $p_0 = p_0^{pent}$ at which $\de$ vanishes.
}
\label{four_cells}
\end{center}
\end{figure}

Equation~\eqref{four_cell_energy} is calculated numerically during a T1 rearrangement (Methods) as shown in Fig.~\ref{four_cells}(a). The total energy during this process is shown in Fig.~\ref{four_cells}(b) as the edge length $\ell$ is contracted (negative values) and a new edge is extended (positive values); the energy barrier $\de$ is the difference in energy between the initial and maximum energy state.  As $p_0$ increases, $\de$ decreases as shown in Fig.~\ref{four_cells}(c). The precise value $p_0^*$ at which energy barriers vanish can be estimated by calculating the energy cost of shrinking an edge of length $\ell= \ell_0$ inside a hexagonal lattice, while all other edges remain unchanged. 
Precisely at the T1 transition, two of the cells are pentagons, while the other two remain hexagonal. Therefore if $p_0< p_0^{pent}= \frac{7+2 \sqrt{7}}{\sqrt{2} \times 3^{3/4}} \approx 3.812$,  pentagons cost finite energy and therefore the transition necessarily requires finite energy . In contrast, for  $p_0 \geq p_0^{pent}$ pentagons (and n-gons with $n>5$) cost no energy and the cells are able to remain in the ground state throughout the transition, requiring zero energy. The estimate $p_0^* = p_0^{pent}$, indicated by a red dashed line in Fig.~\ref{four_cells}(c), does identify the critical target shape index in our mean-field model, and is consistent with the critical point $p_0^* = 3.813\pm 0.005$ identified by the scaling collapse of energy barriers in the full vertex model.

Is there an even simpler explanation for $p_0^* \sim p_0^{pent}$? As in other rigidity transitions~\cite{ LiuNagelReview, Wyart2005, das_quint_schwarz_rigidity,chase_rigidity}, we expect that the critical shape index should also be related to isostaticity. In the vertex model with periodic boundary conditions, cells tile the flat 2D plane, and therefore the total number of vertices $V$, cells $N$, and edges $E$ are related through Euler's formula: $0= V-E+N$. Since each edge is shared by two cells, $E$ is also related to the average coordination number $z$  of cells or $E=N z /2$, which yields $V=N(z/2-1)$. 
The degrees of freedom are simply the motions of each vertex in 2D: $M_{dof} = 2V$.
Assuming force balance (in both directions) and torque balance on each cell generates three constraints per cell: $M_{c} = 3N$. At isostaticity, the number of degrees of freedom equal the number of constraints: $M_{dof} = M_{c}$, resulting in $z_{iso}=5$ and suggesting a mean-field transition at a shape index of $p_0^* \simeq 3.812$. Although it gives a correct prediction, this isostatic argument makes a strong assumption: that constraints are applied to each cell instead of to each vertex.  Therefore, an interesting direction for future research is to understand under what circumstances the energy functional (equation~\ref{scaled_e_tot}) effectively groups vertices into functional units that are cells.

\paragraph{Discussion}

Although the vertex model has been used extensively to model tissues over the past 15 years, there has never been a clear way to connect the model parameters to tissue mechanical properties. Here we show that   
the vertex model has a new and previously unreported critical rigidity transition that occurs at a critical value of the target shape index $p_0^* \sim 3.81$.  This criticality is evident in (a) energy barriers to local T1 rearrangements, (b) the vibrational spectrum of the linear response, and (c) the shear modulus of the tissue. Unlike SPP models where the liquid-to-solid transition is governed by density, our model has a constant-density glass transition governed by single-cell mechanical properties such as cell-cell adhesion and cortical tension encoded in the target shape index $p_0$.

Analyzing only the ground states of the vertex model, the seminal work of Staple et al~\cite{Staple2010} found an ordered-to-disordered transition at $p_0=p_0^{hex} \sim 3.722$. However, because almost all biological tissues are strongly disordered, it remained unclear whether this transition was relevant for the observed glass or jamming transitions in multicellular tissues. Therefore, we explicitly study disordered metastable states and transitions between them.  In addition,~\cite{Staple2010} uses a linear stability analysis of a single cell to suggest that a rigidity transition also occurs at $p_0^{hex}$. In contrast, our analysis explicitly includes multicellular interactions (i.e. collective normal modes) and nonlinear effects (i.e. energy barriers). With this more sophisticated analysis, we demonstrate that vertex models exhibit a rigidity transition at a value $p_0=p_0^{pent} \sim 3.81$ that is measurably different from the prediction $p_0=p_0^{hex}$ based on single-cell linear stability.

Importantly, predictions based on this critical rigidity transition have recently been verified in experiments~\cite{Park_2015}.  Specifically, in both simulations and experiments we can measure the shape index $p = P/\sqrt{A}$ for each cell in a monolayer, where $P$ is the projected cell perimeter and $A$ is the cross-sectional area. In simulations of the vertex model, we find that the median value of the observed shape index $\overline{p}$ is an order parameter that also exhibits critical scaling: $\overline{p} = p_0^* \sim 3.81$ for rigid or jammed tissues and $\overline{p}$ becomes increasingly larger than $p_0^*$ as a tissue becomes increasingly unjammed~Fig.~\ref{scaling_collapse}.  This prediction is precisely realized in cultures from primary cells in human patients, with implications for asthma pathobiology~\cite{Park_2015}. 

We expect that this rigidity framework will help experimentalists develop other testable hypotheses about how the mechanical response of tissues depends on single-cell properties. For example, Sadati \textit{et al.}~\cite{Sadati-Fredberg-review} have proposed a jamming phase diagram where tissues become more \textit{solid-like} as adhesion increases, based on observations of jamming in adhesive particulate matter at densities far below confluency. Using standard interpretations of the vertex model (equations~\eqref{single_cell_energy} and~\eqref{scaled_e_tot}), $p_0$ increases with increasing adhesion, and therefore our model predicts that confluent tissues become more \textit{liquid-like} as adhesion increases. This highlights the fact that adhesion acts differently in particulate and confluent materials; in particulate matter higher adhesion leads to gelation and solidification, while in the vertex model larger adhesion leads to larger perimeters, more degrees of freedom, and liquid-like behavior. These ideas suggest that the role of adhesion in tissue rheology may be much richer and more interesting than previously thought. 

In addition, although all published vertex models assume three-fold coordinated vertices, there is no proof that such structures are stable for $p_0 > p_0^{hex}$~\cite{Staple2010}.  Additionally, higher order vertices are apparently stabilized in some anisotropic biological tissues, including Rosette formation in Drosophila~\cite{Zallen_Karen}.  It will be interesting to study what conditions stabilize higher-fold vertices.

This work may also be relevant to modeling the Epithelial-to-Mesenchymal Transition (EMT) that occurs during cancer tumorigenesis. During EMT, epithelial cells with well-defined, compact shapes and small perimeters relative to their areas transition to mesenchymal cells with irregular shapes and large perimeters relative to their areas~\cite{Kalluri2009}. Since equation~\eqref{scaled_e_tot} specifies a fixed shape index, one could interpret EMT as an increase in $p_0$ leading to a solid-to-liquid transition, providing a simple mechanical explanation for the role EMT plays in metastasis.   In order to explore this idea further, it will be necessary to determine if a similar rigidity transition exists in three dimensions. A simple extension of this model would replace perimeters and areas in Eq.~\ref{scaled_e_tot} with surface areas and volumes, respectively; this is a promising avenue for future work. 

We expect that this model may be of interest to scientists independent of its biological relevance.  We have shown it exhibits a simple rigidity transition with a novel control parameter, and therefore it might provide a useful bridge between jamming transitions in particulate matter~\cite{Wyart2005, LiuNagelReview} and rigidity transitions in random elastic networks~\cite{das_quint_schwarz_rigidity,chase_rigidity, frey}. In particular, the potential grouping of vertices into functional cell units could draw an explicit connection between spring networks and particle/cell packings. An open question is whether our model belongs to an existing universality class, and whether the transition is mean-field.

Finally, the fact that the vertex model exhibits disordered ground states for $p_0 > p_0^*$ suggests that it may be a useful toy model for thermodynamic (as opposed to kinetic) explanations of the glass transition in particulate matter. Furthermore, these states are predicted to be hyperuniform~\cite{GabrielliTorquato} with a photonic band gap, indicating that they may be useful for designing metamaterials with interesting optical properties.

\section{Methods}
\subsection{Simulating a confluent tissue monolayer}
 To simulate confluent monolayers, a Random Sequential Addition point pattern~\cite{RSA_Torquato} of $N$ points was generated under periodic boundary conditions, with box size $L$ chosen such that the average area per cell is unity. Two methods of generating this initial point pattern were used: a Random Sequential Addition point pattern \cite{RSA_Torquato}, and a Poisson point pattern.  The results presented in this work are independent of the method of initial point pattern generation.  
A voronoi tessellation of this point pattern results in a disordered cellular structure, which was then used as an input to the program \textit{Surface Evolver}~\cite{Brakke}.  Surface Evolver numerically minimizes the total energy of the system (equation~\eqref{scaled_e_tot}) at fixed topology using gradient descent with respect to the vertices of the cells. If an edge shrinks below a threshold value $l^*$, a passive T1 transition is allowed if it lowers the energy. All structures are minimized such that the average energy of a cell changes by less than one part in $10^{10}$ between consecutive minimization steps, and as in other simulations of the vertex model~\cite{Hufnagel2007, Staple2010}.

Once an initial energy-minimized state is reached, T1 transitions are actively induced at every edge to measure energy barriers~\cite{bi_softmatter}.  An example of a T1 in the simple four-cell case is shown in Fig.~\ref{de_stat}(a): the central thick edge is quasi-statically contracted to zero length ($\ell=0$) at which point a T1 topological swap is executed.  After the T1, the length of the central edge is then expanded until it reaches the initial length ($\ell= \ell_0$). The total energy of four cells during this process is shown in Fig.~\ref{four_cells}(b); the edge length is represented by a negative value during contraction and flips sign after the T1.  

For each active T1 transition in an N-cell system, the energy barrier is defined as the total energy difference between the initial state $\ell=\ell_0$ and the onset of T1 topological swap ($\ell=0$). Calculations of energy barriers were repeated for various values of $r$ at decade increments from $0.005$ to $200$ and $p_0$ ranging from $3$ to $4$.

To calculate the shear modulus, we apply quasistatic simple shear to a tissue using Lee-Edwards periodic boundary conditions. The shear modulus is calculated by taking the linear response of the tissue, 
\begin{equation}
G_{xy} = \frac{1}{L^2} \lim_{\gamma_{xy} \to 0} \frac{\partial^2 \varepsilon}{\partial \gamma_{xy}^2},
\end{equation}
where $L$ is the linear dimension of the tissue. The results for $G_{xy}$ at each value of $r$ and $p_0$ were obtained by averaging 20 runs. 
 
\subsection{Calculation of the vibration density of states}
We obtain the vibrational density of states by diagonalizing the Hessian matrix of the system
\begin{equation}
H_{i \mu j \nu} = \frac{\partial^2 \varepsilon}{\partial r_{i\mu} r_{j\nu}},
\label{hessian}
\end{equation}
where $i,j$ are indices for vertices and $\mu, \nu$ cartesian coordinates, and $\varepsilon$ is defined in equation~\eqref{scaled_e_tot}. The eigenvalues of equation~\eqref{hessian} are  $\{ \lambda_i \}$.

\section{Acknowledgements} 
We would like to thank M. C. Marchetti for useful comments on this manuscript. M.L.M. acknowledges support from the Alfred P. Sloan Foundation, and M.L.M and D.B. acknowledge support from NSF-BMMB-1334611 and NSF-DMR-1352184. M.L.M and D.B. also would like to thank the KITP at the University of California Santa Barbara for hospitality and was supported in part by the National Science Foundation under Grant No. NSF PHY11-25915.


\begin{thebibliography}{48}
\bibitem [{Note2()}]{note_p0}%
  \BibitemOpen
  \bibinfo {note} {In this Letter we focus on the regime where $p_0 > 0$, but
  $p_0 < 0$ is a valid parameter regime where our results hold as
  well.}\BibitemShut {Stop}%
\end{thebibliography}

%
\section{SUPPLEMENTARY INFORMATION: Effect of variations in single cell properties}
\begin{figure}[htpb]
\begin{center}
\includegraphics[width=1\columnwidth]{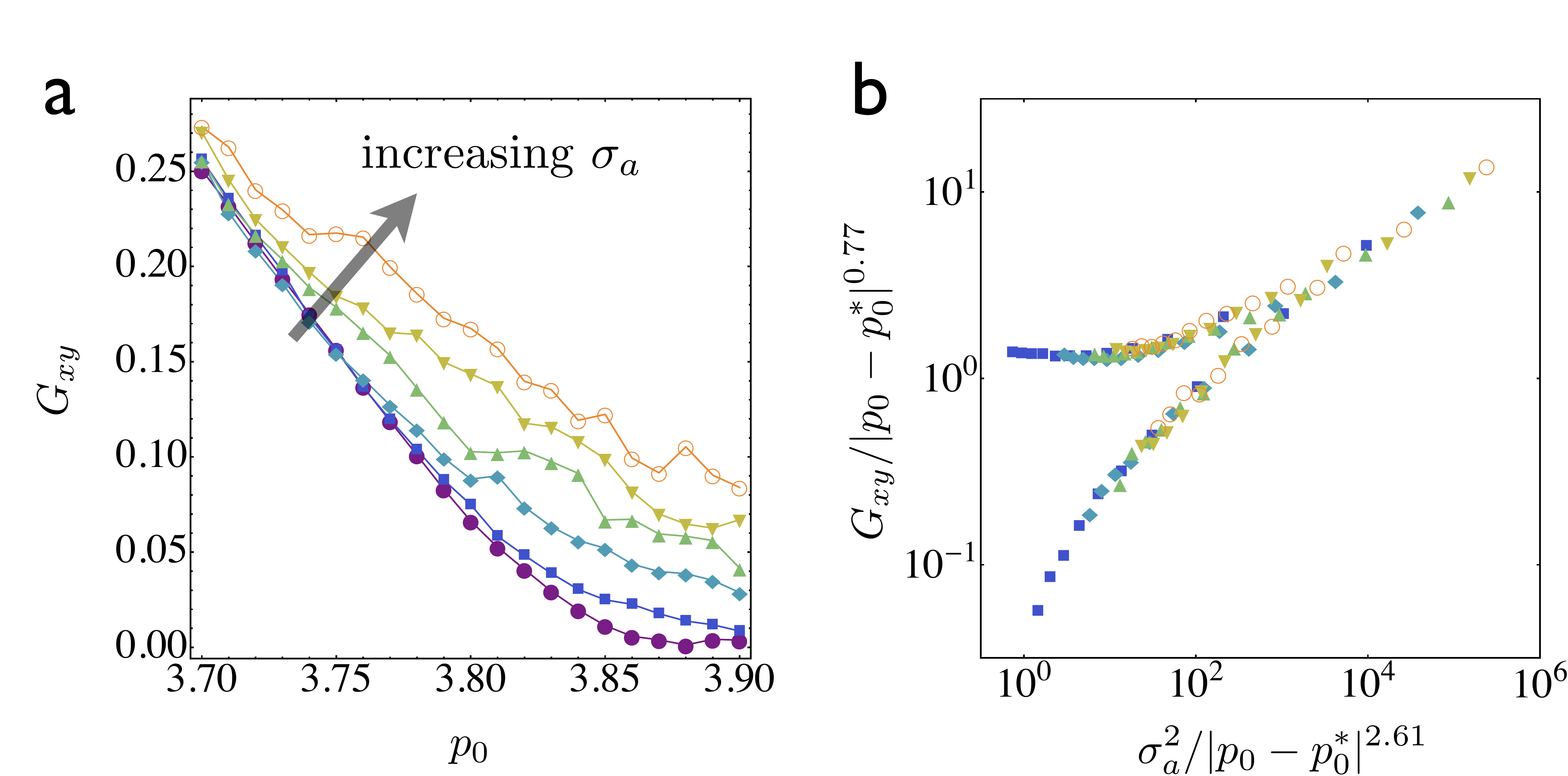}
\caption{
The shear modulus of a tissue with varying preferred cell area. 
(a) The shear modulus as function of $p_0$ for different values of $\sigma_a= 0,~0.05,~0.1,~0.15,~0.2~\& ~0.25$ (from bottom to top).  
(b) The shear modulus obeys a universal scaling function. The location of the rigidity transition is found to be $p_0^*=3.813$. 
}
\label{vary_a}
\end{center}
\end{figure}

To study the effect of non-homogeneous cellular properties on the rigidity transition, we first probe the mechanical property of a tissue with a randomly distributed `preferred' cell area
\begin{equation}
\label{scaled_e_tot}
\mathcal{\epsilon} = \sum_{i} \left [ (\tilde{a}_i-a_0)^2 + \frac{(\tilde{p}_i - p_0)^2}{r} \right],
\end{equation}
where $a_0$ is drawn from a Gaussian distribution with mean $1$ and variance $\sigma_a^2$. We have chosen to use $r=0$ for simplicity, which does not affect the result. The shear modulus is plotted as function of $p_0$ in Fig.~\ref{vary_a}(a) for various values of $\sigma_a$. The data at $\sigma_a=0$ correspond to that shown in the main manuscript. $\sigma_a$ introduces more disorder in cell areas and results in more fluctuations near the rigidity transition. As $\sigma_a$ is increased,  the rigidity transition is `softened' similar to the effect of increasing $r$. A small value $\sigma_a$ of should provide a white noise or thermal-like fluctuation to the tissue and we hypothesize the scaling form for the shear modulus. 
 $ G_{xy}  / \lvert p_0-p_0^*\rvert^{\phi_a} \propto  \sigma_{p_0}^2 / \lvert p_0-p_0^*\rvert^{\Delta_a}$.
Fig.~\ref{vary_a}(b) shows that the data from Fig.~\ref{vary_a}(a) can be scaled to collapse using  $\Delta_a=2.6$, $\phi_a=0.77$ and at the same value of $p_0^*=3.813$. This suggests that adding disorder in preferred cell areas does not change the location of the rigidity transition. 

\begin{figure}[htpb]
\begin{center}
\includegraphics[width=1\columnwidth]{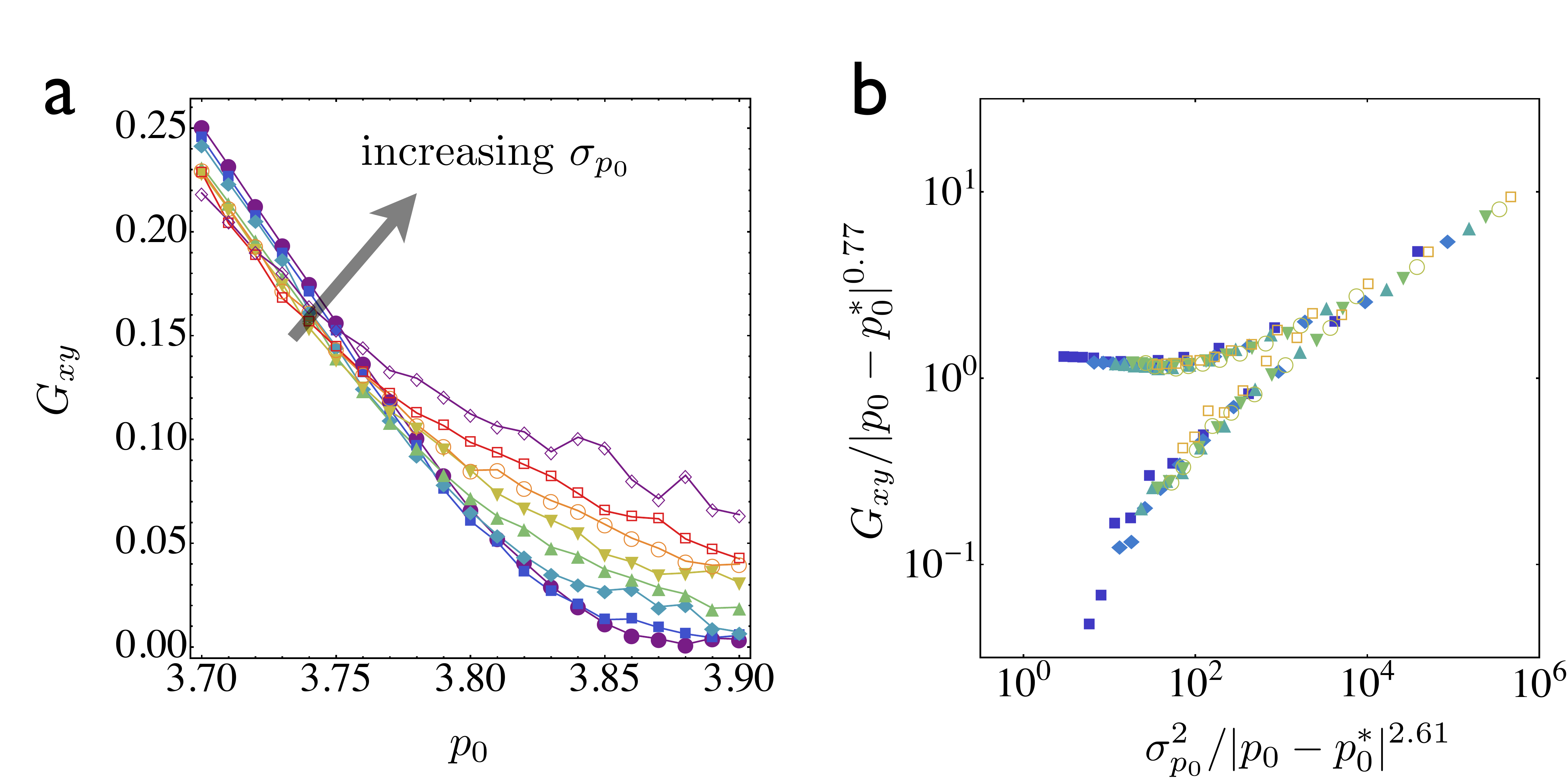}
\caption{
The shear modulus of a tissue with varying $p_0$. 
(a) The shear modulus as function of $p_0$ for different values of $\sigma_a=0,~0.05,~0.15,~0.2,~0.25,~0.3,~0.35,~0.4$(from bottom to top).  
(b) The shear modulus obeys a universal scaling function. The location of the rigidity transition is found to be $p_0^*=3.813$. 
}
\label{vary_p}
\end{center}
\end{figure}

A similar numerical calculation was performed for tissues with varying value of $p_0$. With mean of $p_0$ and variance $\sigma_{p_0}^2$. The same analysis can be carried out to show that the location of the rigidity transition does not shift with  $\sigma_{p_0}^2$ (Fig.~\ref{vary_p}).

\end{document}